\documentclass[10pt,journal]{IEEEtran}
\makeatletter
\def\ps@headings{%
\def\@oddhead{\mbox{}\scriptsize\rightmark \hfil \thepage}%
\def\@evenhead{\scriptsize\thepage \hfil \leftmark\mbox{}}%
\def\@oddfoot{}%
\def\@evenfoot{}}
\makeatother 
\usepackage[dvips]{epsfig}
\usepackage[dvips]{graphicx}
\usepackage{latexsym}
\usepackage{amssymb,balance}
\usepackage{graphicx}
\usepackage{subfigure}
\usepackage{subfig}
\begin{document}
\title{- \emph{Technical Report} -\\ Beaconless Geo-Routing Under The Spotlight: Practical Link Models and Application Scenarios}
\author{Ahmed Bader, Karim Abed-Meraim, \emph{Senior Member}, \emph{IEEE}, and Mohamed-Slim Alouini, \emph{Fellow}, \emph{IEEE}


}


\maketitle

\begin{abstract}
Analysis and simulation of beaconless geo-routing protocols have
been traditionally conducted assuming equal communication ranges for
the data and control packets. In reality, this is not true since the
communication range is actually function of the packet length.
Control packets are typically much shorter than data packets. As a
consequence, a substantial discrepancy exists in practice between
their respective communication ranges. In this paper, we devise a
practical link model for computing the effective communication
range. We further introduce two simple strategies for bridging the
gap between the control and data packet communication ranges. Our
primary objective in this paper is to construct a realistic
analytical framework describing the end-to-end performance of
beaconless geo-routing protocols. Two flagship protocols are
selected in this paper for further investigation under the developed
framework. For a better perspective, the two protocols are actually
compared to a hypothetical limit case; one which offers optimal
energy and latency performance. Finally, we present four different
application scenarios. For each scenario, we highlight the
geo-routing protocol which performs the best and discuss the reasons
behind it.
\end{abstract}



\IEEEpeerreviewmaketitle


\section{Introduction}\label{sec:intro}
Beaconless geo-routing protocols have emerged as some of the most
efficient packet delivery solutions for Wireless Sensor Networks
(WSNs) \cite{wsn.survey.Yicka,beaconless.comparison.Sanchez} as well
as Mobile Ad Hoc Networks (MANETs) \cite{hybrid.beaconless.Xiang}.
This is mainly due to the fact that nodes can locally make their
forwarding decisions using very limited knowledge of the overall
network topology. This comes very handy for mobile applications as
well as for scenarios where random sleeping schedules are applied.
Consequently, beaconless geo-routing offered substantial enhancement
in terms of bandwidth efficiency in comparison to their beacon-based
predecessors \cite{beaconless.comparison.Sanchez}.\\
\indent In beaconless geo-routing, potential relays must undergo
first a selection process whereby the node with the most favorable
attributes (e.g. closeness to destination) shall eventually forward
the packet \cite{beaconless.comparison.Sanchez}. The selection
process is triggered by the sender using a Request-To-Send (RTS)
message. The key concept here is to weigh the response time of
potential relays according to their forwarding attributes. Two of
the earliest such protocols reported in literature are Geographic
Random Forwarding (GeRaF)
\cite{GeRaF.latency.zorzi,GeRaF.multihop.zorzi} and Beaconless
Routing (BLR) \cite{BLR.Heissen}. In GeRaF, the relay selection
process is controlled by the packet sender at any given hop. The aim
is to select the node lying within the communication range and which
is closest to the destination. In BLR, the relay selection criteria
is very similar but the process itself is rather distributed. The
ideas presented in \cite{GeRaF.latency.zorzi,GeRaF.multihop.zorzi}
and \cite{BLR.Heissen} are believed to have fueled the research in
this area over the decade to follow. Many of the geo-routing
protocols have embraced on the key concepts presented in those early
works. For instance, in Contention-Based Forwarding (CBF)
\cite{contention.based.Fussler}, response time to the RTS message is
rather calculated as function of the advancement offered by a
candidate relay towards the destination. On the other hand, the
response time of potential relays in MACRO \cite{MACRO.Ferrara} is
weighed by the progress that can be made per unit power. In Cost and
Collision Minimizing Routing (CCMR) \cite{cost.minimizing.Rossi},
the authors propose a technique whereby contending relays
dynamically adjust their cost metrics during the selection process.
GeRaF itself is modified in M-GeRaF
\cite{M-GeRaF.performance.Odorizzi}, such that it serves wireless
sensor networks with multiple sinks. Implicit Geographic Forwarding
(IGF) \cite{IGF.He} proposes two optimizations to reduce the number
of responses and collisions during the relay selection process. Yet,
one of the most noteworthy twists in beaconless geo-routing was
offered in Beaconless On Demand Strategy for Geographic Routing
(BOSS)\cite{BOSS.Sanchez}. In this protocol, the sender piggybacks
the data payload in the RTS control packet. This was mainly
motivated by the discrepancy in the communication ranges between the
data packet (containing the payload) and the control packet. Another
recent and interesting addition to the geo-routing protocol family
is CoopGeo \cite{CoopGeo.Aguilar} which combines cooperative
relaying and beaconless geo-routing with the objective of enlarging
the progress made every hop. Finally, a hybrid mechanism has been
recently devised in \cite{hybrid.beaconless.Xiang} to switch between
beacon-based and beaconless states based on the
underlying application scenario.\\
\indent Despite the breadth of development in the beaconless
geo-routing protocol family, analysis and simulation of these
protocols have been carried out often using simplistic link models.
To be more specific, the effect of the packet length on the
probability of successful packet detection has not been considered
at all except in BOSS \cite{BOSS.Sanchez}. Empirical test results
obtained by the authors in \cite{BOSS.Sanchez} indicated that the
average packet error rate (PER) for the data message is notably
higher than that of the control message. This is true simply due to
the fact that the data packet is typically much larger in size. In
fact, this is expected and is inline with literature
\cite{conv-coding.PER.Awoyini}. When convolutional coding is
utilized (which is often the case), the average PER grows with the
packet length. The growth in PER is associated with larger signal to
noise and interference ratio (SINR) targets and therefore shorted
communication ranges.\\
\indent Another major area where improvement is deemed essential
relates to the assessment of end-to-end performance. The vast
majority of beaconless geo-routing protocols have been investigated
only from the perspective of a single node or a single hop.
End-to-end performance has been seldom considered, and whenever
considered, it has been studied using empirical test or simulations.
Results obtained from such approaches are valuable indeed. However,
they are limited to a finite set of scenarios and parameter values.
Consequently, we believe that it is quite instrumental to develop an
analytical framework for the evaluation of the end-to-end
performance. Such a framework should offer the research community a
readily available tool to study as well as optimize the performance
of beaconless geo-routing protocols under any arbitrary choice of
scenarios and
parameters.\\
\indent Based on the above, it is our primary objective in this
paper to develop an end-to-end analytical framework for beaconless
geo-routing protocols while taking into consideration a more
practical link model. In doing so, we do not analyze all geo-routing
protocols as this would require a substantial amount of work.
Instead, we have reverted to studying two representative protocols:
GeRaF and BOSS. A detailed rationale behind this selection is
provided in Section \ref{sec:assess}. Subsequently, our main
contribution in this paper is twofold:
\begin{enumerate}
\item Providing an end-to-end analytical framework for two prominent
beaconless geo-routing protocols. This framework can be conveniently
adapted to other protocols within the same family. \item
Incorporating practical link models in the analysis which take into
consideration fading as well as the effect of packet length on the
average PER.
\end{enumerate}
The paper is organized as follows. Section \ref{sec:overview}
provides an overview of beaconless geo-routing protocols. In
specific, Section \ref{sec:assess} rationalizes the selection of
BOSS and GeRaF as the focus of our study. Section \ref{section:PHY}
presents the wireless channel as well as the wireless link models.
The end-to-end performance of BOSS and GeRaF is analyzed in depth in
Section \ref{sec:e2e_analysis}. Four distinct application scenarios
are investigated in the context of beaconless geo-routing in Section
\ref{sec:eval}. Finally, key examples on how to optimize end-to-end
performance of beaconless geo-routing are illustrated in Section
\ref{sec:opt}. A summary of notations used in this paper is provided
in Table \ref{table:notations}.
\begin{table}[t]
\caption{Table of Notations}
\begin{center}
\begin{tabular}{|l|l|}

\hline\hline

$D$     &       Distance between source and destination\\
$L$     &       Packet length in bits\\
$P_{t_{max}}$   &   Maximum transmit power available for a node\\
                &   (limited by hardware and energy constraints)\\
$P_{t_D}$   &   Transmit power for data messages (GeRaF)\\
& and RTS messages (BOSS)\\
$P_{t_C}$   &   Transmit power for control packets\\
$P_{t_{BT}}$   &    Transmit power on the busy tone\\
$P_{Rx}$    &       Power consumed when in receive state\\
$P_n$       &       Noise power\\
$a_i$       &       Area of the $i$th forwarding slice,
$i=1,\ldots,2N$\\
$T_p$       &       Data packet duration\\
$T_c$       &       Minimum control packet duration\\
$T_s$       &       Duration of a forwarding slot\\
$N$         &       Number of forwarding subareas\\
$n_T$       &       Number of data packet transmissions (GeRaF-MRC)\\
$\gamma$    &       Instantaneous SINR\\
$\gamma_{t_C}$&      Detection threshold, control packet\\
$\gamma_{t_D}$&      Detection threshold, data packet\\
$h_n$       &       Fading coefficient of the $n$th multipath\\
$R$         &       Communication range\\
$\lambda$   &       Wavelength\\
$\alpha$    &       Path loss coefficient\\
$\rho$      &       Network node density\\
$\epsilon$  &       Sleeping duty cycle\\
$\eta$      &       Number of cycles elapsing without a CTS
response\\
$m_e$       &       Number of slots elapsing without a CTS
response\\
$m_n$       &       Number of slots required to resolve a
collision\\
$x$         &       Random offset from the beginning of the control slot\\
& (applicable to BOSS)\\
PPA         &       Positive progress area\\
NPA         &       Negative progress area\\

 \hline\hline

\end{tabular}
\end{center}
 \label{table:notations}
\end{table}

\section{Beaconless Geo-Routing Overview}\label{sec:overview}
In broad terms, beaconless geo-routing protocols operate as per the
following guidelines. The sender's communication range is divided
into two areas. The first area is the one offering positive progress
towards the destination and is denoted as PPA. In other word, relays
lying in the PPA are closer to the destination than the sender. The
complementary of this area is the negative progress area (NPA). Each
area is further sliced into $N$ forwarding subareas. Two interesting
alternative for slicing PPA and NPA are illustrated in
\cite{CoopGeo.Aguilar}(Fig. 2) and \cite{BOSS.Sanchez}(Fig. 3). The
sender of a packet first issues a request-to-send (RTS) message.
Upon the reception of this message, potential relays lying within
the sender's coverage zone enter into a time-based contention phase.
Some protocols such as GeRaF, BLR, IGF, CBF, MACRO, and CCMR exclude
nodes in the NPA right away. Others such as BOSS and CoopGeo may
revert to those nodes at a later stage in the relay selection
process. Each potential relay triggers a timer whose expiry depends
on a certain cost function. The first node to have its timer expire
will transmit a clear-to-send (CTS) message on the next available
time slot. However, since time is slotted it is probable that
collisions may occur. A secondary collision resolution phase may be
devised in this case.

\subsection{Assessment of Various Protocols}\label{sec:assess}
As mentioned in Section \ref{sec:intro}, we have selected GeRaF and
BOSS as a baseline for our study. The selection of GeRaF stems from
the fact that its performance on per-node basis is well understood
and elaborately analyzed in \cite{GeRaF.latency.zorzi} and
\cite{GeRaF.multihop.zorzi}. On the other hand, BOSS has been shown
in practice to excel in certain aspects of performance
\cite{beaconless.comparison.Sanchez}. Justifications for not
including other protocols in this study are summarized in the
following.
\begin{enumerate}
\item BLR was not studied further for the simple reason that it has been observed to suffer from
frequent packet duplications and collisions
\cite{beaconless.comparison.Sanchez,optim.beacons.Heissenbuttel}.
\item CBF is also expected to suffer to some extent from the same problem as BLR
since the relay selection is carried out in a distributed fashion.
CBF proposes to reduce the impact of packet duplication by means of
devising a suppression phase. However, this is expected to drive the
protocol to consume more energy and produce larger forwarding
delays. \item MACRO is based on utilizing the residual energy as a
relay selection metric. It is shown in \cite{MACRO.Ferrara} that
MACRO outperforms GeRaF \emph{only} slightly in certain aspects of
single-hop performance. Furthermore, our preliminary investigations
revealed that the end-to-end delay performance of both protocols
tends to be quite comparable. \item M-GeRaF is a mult-sink extension
of GeRaF. It may be useful in infrastructure-based applications with
multiple sinks or for multicast applications. The scope of this
paper is restricted however to unicast scenarios. \item On the other
hand, a deeper look at IGF exposes two aspects which may jeopardize
its ability to perform well. The forwarding region which contains
candidate relays is restricted in IGF to a small sector
\cite{IGF.He}(Fig. 2). The rationale behind this is to increase the
probability that all candidate relays lie within the communication
range of each other. As such, it is assumed that collisions are very
unlikely to occur. In practice, restricting the size of the
forwarding region limits the average number of candidate relays to a
small subset. This may result in empty cycles wherein there will no
candidate relays. This needs to be taken into consideration for the
sake of an objective analysis. Furthermore, IGF assumes that in the
case of collisions, the sender has the ability to resolve duplicate
CTS responses by choosing only one. However, authors do not indicate
how this is achieved. The lack of an explicit mechanism for
resolving colliding packets only leaves room for speculations about
how IGF would perform. Based on the above reasoning we have selected
not to pursue IGF any further.
\item CoopGeo on the other hand focuses on creating diversity during
the transmission process by selecting two relays instead of one. The
primary relay is the one that maximizes progress towards the
destination. The secondary one offers diversity. In that sense,
CoopGeo has been designed in a way to reduce PER without really
considering its impact on energy performance. The study has been
also limited to a single-hop case. Although we do acknowledge the
value of contributions of CoopGeo, we rather believe it should be
deferred as a subject of future research. \item At the other end of
the spectrum, CCMR indeed promises performance levels which are
superior to GeRaF. However, the dynamic nature of the relay
selection process makes it nothing but straightforward to derive
meaningful expressions for the end-to-end performance.
\end{enumerate}
It is essential to note here that we are not overlooking the
performance bounds that CCMR is poised to achieve. In order to put
things into the right perspective, we have introduced a hypothetical
beaconless geo-routing protocol which is able to achieve the optimum
end-to-end performance. Although we do not know for a fact where
CCMR stands with respect to this optimum performance level, we will
have the chance now to compare GeRaF and BOSS against the
performance limits of beaconless geo-routing. In the context of
beaconless geo-routing, the optimum protocol is evidently the one
which involves the minimum number of transactions during the relay
selection phase. This simply translates to one RTS message from the
sender, one CTS response from the best candidate relay, followed by
packet transfer. The optimum forwarding process here is denoted as
RTS/CTS/DATA-opt.

\subsection{Bridging The Gap}
As mentioned in Section \ref{sec:intro}, BOSS has been originally
designed such that the data payload is incorporated inside the first
RTS message. By doing that, only those nodes who are able to
successfully receive the data packet will contend for becoming the
next forwarder \cite{BOSS.Sanchez}. However, transmitting the data
and control packets at the same maximum available power obviously
results in a range gap. In other words, there will be some nodes who
were not able to receive the RTS message successfully but rather
will be able to receive the subsequent control packets. This is
clearly a waste of node energy. Control packets can have the same
range as the data packet while using lower transmit power. To
alleviate this shortcoming, we assume that nodes transmit at a lower
power level when sending a control message. To offer a fair
comparison, GeRaF is equipped with the same capability. In other
words, nodes are able to reduce the transmit power on the control
packet to the level that the resulting range matches that of the
data packet. GeRaF in this case is labeled as GeRaF-PC. In the
specific case of GeRaF however, there is more than one way to bridge
the gap between the data and control packets. One method that we
have studied here is to utilize time diversity with maximal ratio
combining (MRC) at the receiver. Under this scheme, the sender
transmits control and data packets at the maximum power. The data
packet is retransmitted as many times as needed to make the two
communication ranges equivalent. This version of GeRaF is thus
tagged as GeRaF-MRC. The average number of transmissions needed is
denoted as $n_T$.

\section{Wireless Link Model}\label{section:PHY}
In this section, we first highlight some key elements to be
considered with respect to the underlying wireless channel model. We
then shed some light on how to derive a numerical relationship
between packet length and average PER. This is instrumental to lay
down a practical model for establishing a successful link between
two nodes.

\subsection{Key Considerations}
\begin{enumerate}
\item Unpunctured convolution coding is utilized with code rate
$\frac{1}{2}$. Hard-decision decoding is assumed. Unless explicitly
mentioned, QPSK modulation is used.
\item The physical layer (PHY) is loosely based on the widely adopted IEEE
802.15.4-2006 standard \cite{WSN.Zigbee.Baronti}. \item Nodes are
generally mobile thus the channel is time-selective with respect to
the packet but is assumed to be constant within the symbol duration.
For application scenarios where nodes are stationary, we assume a
quasi-static fading channel.
\item For WSN applications, we may conveniently assume that the
channel is frequency non-selective. However, for some MANET
applications such as Vehicular Ad Hoc Networks (VANETs), the channel
is indeed frequency selective.
\end{enumerate}
The time and frequency selectivity of the channel requires further
elaboration. For WSN application scenarios, the communication ranges
are typically short; in the range of a couple of hundred meters.
Consequently, the delay spread of the wireless channel is also
small. As such, the fading channel can be assumed to be frequency
non-selective. The occupied bandwidth in IEEE 802.15.4-2006 is $2$
MHz \cite{ZigBee.BW.Ruben}. For the channel to be otherwise
frequency-selective, the delay spread should be greater than
$0.5\mu$s. This corresponds to an excess path length of $150$
meters. Communication ranges in WSN applications are relatively
short such that an excess path length great than $150$ meters is
quite unlikely. Furthermore, nodes are typically stationary in WSNs
thus the channel can be assumed to be
quasi-static as mentioned above.\\
\indent On the other hand, nodes in VANET applications enjoy better
access to energy resources, and thus are able to transmit at higher
power levels. As such, communication ranges are relatively larger
and so are the excess path lengths. This drives the channel towards
becoming frequency selective. As a consequence, this mandates nodes
to utilize some sort of channel equalization, e.g. a zero-forcing
equalizer which in essence performs as a RAKE receiver. From a
time-domain perspective, the channel varies over the duration of a
single packet. This definitely needs to be taken into consideration
when developing the packet detection model.

\subsection{Dependency of Range on Packet Length}
\begin{figure}[t]
\begin{center}
\epsfxsize=7cm \centerline{\epsffile{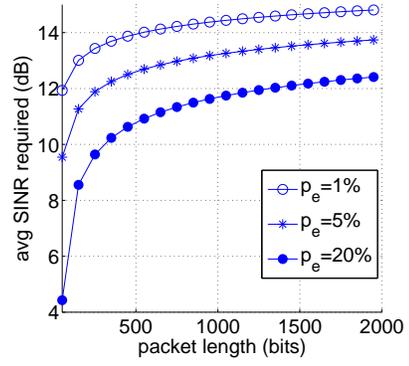}}
\end{center}
\caption{Relationship between the required average SINR and the
packet length for a range of PER targets.} \label{fig:a_avg_Lcc}
\end{figure}

This subsection outlines the method for deriving a relationship
between the length of the packet and the detection threshold. In
\cite{conv-coding.PER.Awoyini}, the authors derive an expression for
the PER of an Orthogonal Frequency Division Multiplexing (OFDM)
system utilizing convolutional coding. The channel in
\cite{conv-coding.PER.Awoyini} is assumed to be quasi-static and
frequency selective. Thus, the channel vector $\mathbf{H}$
represents a set of uncorrelated fading coefficients corresponding
to the OFDM subcarriers. In our case, the channel is assumed to be
frequency non-selective or is forced to become so by means of a RAKE
equalizer. In the case of stationary nodes, the channel simply
becomes a scalar. However, for mobile nodes, the channel is actually
time-varying. Nevertheless, we can still utilize the same analytical
framework in \cite{conv-coding.PER.Awoyini} here. However,
$\mathbf{H}$ would be now representing the channel
fading coefficients over time instead of frequency.\\
\indent In our case, time is divided into equal blocks whereby the
duration of one block equals the channel coherence time. Obviously,
the channel coherence time depends on the level of node mobility and
may be computed using \cite{wireless.book.Rappaport} (Eq. 4.40.c).
To proceed, we assume that the channel fading coefficient within
each coherence block is constant and is uncorrelated with respect to
the other blocks. For the sake of simplification, we assume that the
coherence time is an integer multiple of the symbol duration. The
SINR in a given coherence block is denoted by $\gamma$. As will be
discussed in the next subsection, $\gamma$ is exponentially
distributed with mean $\overline{\gamma}$. Based on the above, we
can now utilize the analysis offered in
\cite{conv-coding.PER.Awoyini} to derive a numerical relationship
between the length of the packet and $\overline{\gamma}$.
This is plotted in Figure \ref{fig:a_avg_Lcc} for various PER
targets. It is worthwhile to note that for loose PER targets (e.g.
20\%), a 20 fold increase in packet length resulted in an 8-dB
growth in the required SINR. This growth is less drastic in case of
lower PER targets.

\subsection{Packet Detection Criteria}
The SINR threshold required to successfully receive a control packet
and a data packet are denoted by $\gamma_{t_C}$ and $\gamma_{t_D}$
respectively. As the length of the control message is shorter in
length than the data message, then $\gamma_{t_C}<\gamma_{t_D}$. The
fading coefficients are assumed to be complex Gaussian. As a generic
case, the channel is considered to be frequency selective. It is
modeled by a tap-delay line with coefficients $\{h_n\}_{n=1}^m$,
where $m$ is the number of multipaths
\cite{wireless.book.Rappaport}. In case of a frequency non-selective
channel $m=1$. The SINR over an arbitrary communication link is
given by $\gamma=\frac{P_t}{P_n}\sum_{n=1}^m|h_n|^2$, where $P_t$ is
the transmit power and $P_n$ is the noise floor. The SINR $\gamma$
is exponentially distributed with mean
$\overline{\gamma}=\frac{2P_t}{P_n}\sum_{n=1}^m\mathbb{E}[|h_n|^2]$,
where $\sum_{n=1}^m\mathbb{E}[|h_n|^2]$ is function of the distance
between the transmitter and the receiver. A control packet is
successfully detected if $\overline{\gamma}\geq\gamma_{t_C}$.
Similarly, $\overline{\gamma}\geq\gamma_{t_D}$ is the condition for
successful detection of the data message. Denoting the communication
range by $R$, we have
$\sum_{n=1}^m\mathbb{E}[|h_n|^2]=\left(\frac{\lambda}{4\pi
R^2}\right)^\alpha$ at the edge of the range, where $\alpha$ is the
path loss exponent and $\lambda$ is the wavelength. Consequently,
the communication range for the data message is expressed as
\begin{equation}
R=\sqrt{\left(\frac{\lambda}{4\pi}\right)\left(\frac{2P_{t_D}}{\gamma_{t_D}P_n}\right)^{1/\alpha}}\label{eq:R},
\end{equation}
where $P_{t_D}$ is the transmit power of the data packet. The range
for the control packet is obviously obtained by
substituting $\gamma_{t_D}$ and $P_{t_D}$ in (\ref{eq:R}) with $\gamma_{t_C}$ and $P_{t_C}$ respectively.\\
\indent For GeRaF-MRC, we assume i.i.d. channel fading coefficients
every time the packet is transmitted. The desired MRC gain here is
$\gamma_{t_D}/\gamma_{t_C}$. As such, the number of packet
transmissions required is $n_T=\lceil \gamma_{t_D}/\gamma_{t_C}
\rceil$.

\section{End-to-End Performance Analysis}\label{sec:e2e_analysis}
Nodes are assumed to be distributed according to a Poisson Point
Process (PPP) with an average density of $\rho$. Nodes are assumed
to employ asynchronous sleeping schedules with a duty cycle of
$\epsilon$. Furthermore, nodes are assumed to activate a busy tone
(BT) during listening and receiving to help mitigate the hidden node
effect \cite{GeRaF.latency.zorzi,dualBTMA.adhoc.Haas}. Originally,
BOSS neither incorporates sleeping schedules nor a busy tone.
Nevertheless, BOSS is equipped here with these tools for the sake of
a fair and objective study. The duration of the data and control
messages are denoted by $T_p$ and $T_c$ respectively. It is assumed
in this paper that both versions of GeRaF as well as BOSS are
utilizing the smallest possible packet size for the control message.
Control packets are transmitted at a power level of $P_{t_C}$ while
data packets are transmitted at $P_{t_D}$. The power consumed while
in receive state is $P_{Rx}$ while the transmit power for the BT is
$P_{t_{BT}}$. In the next two subsections, we offer a detailed
analysis of the end-to-end performance of GeRaF-PC, GeRaF-MRC, and
BOSS.

\subsection{GeRaF}
The forwarding process occurs in time over successive cycles. One
cycle consists of $N$ time slots corresponding to the number of
forwarding subareas in PPA. The duration of one slot is $T_s$. GeRaF
is designed such that each slot consists of two parts. The first is
always reserved to the sender while the second contains responses
from candidate relays. As such, $T_s=2T_c$. There are three main
types of messages that the sender may send over the first half of
the control slot: RTS, CONTINUE, and OK. The CONTINUE message
indicates the occurrence of a collision and triggers a new round of
contention between the relays. Collisions typically occurs between
those relays offering the best progress towards the destination,
i.e. those lying in the foremost forwarding subarea. The OK message
simply informs the successful relay of being selected as the
next-hop forwarder.\\
\indent At any given hop, there would be $\eta$ empty cycles
followed by one non-empty cycle. Empty cycles occur when there are
no awaken nodes in the PPA. In the non-empty cycle, there would be
$m_e$ empty slots followed by $m_n$ collision-resolution slots. The
$m_e$ empty slots reflect the fact that there are no awaken nodes in
the first $m_e$ forwarding subareas of the PPA. The expectations
$\mathbb{E}[\eta]$, $\mathbb{E}[m_e]$, and $\mathbb{E}[m_n]$ are
found in explicit forms in \cite{GeRaF.latency.zorzi} ((3) and (4)).
Table \ref{table:GeRaF-MRC} provides a description of the relay
selection and packet forwarding process during a given hop of
GeRaF-MRC. For each activity, Table \ref{table:GeRaF-MRC} indicates:
\begin{enumerate}
\item The duration of the activity, $t_y$. \item The nature of the nodes involved in
the activity. \item The average count of nodes involved, $n_y$.
\item The task associated with this activity. \item The power factor
at which the associated task is conducted, $pf_y$.
\end{enumerate}
It is important to note that activities may overlap in time. For
instance activities 3 and 4 in Table \ref{table:GeRaF-MRC} take
place at exactly the same time: the sender issues a RTS message
while awaken nodes react to it.\\
\indent The energy consumed to accomplish a given activity is
$t_yn_ypf_y$. The overall energy consumed per hop, $E_{hop}$ is the
sum of all individual transmission and reception activities
undertaken to select the successful relay and then send the packet
to it. As per Table \ref{table:GeRaF-MRC}, groups of activities are
repeated multiple times. For example, activities 1 and 2 in Table
\ref{table:GeRaF-MRC} are repeated on average $\mathbb{E}[\eta]$
times. For GeRaF-MRC, $P_{t_D}=P_{t_C}=P_{t_{max}}$ which is the
maximum transmit power available to a node. We also note that for
GeRaF-PC, $P_{t_C}=P_{t_D}\frac{\gamma_ {t_C}}{\gamma_{t_D}}$, while
$P_{t_D}=P_{t_{max}}$. On the other hand, it can be shown in light
of Table \ref{table:GeRaF-MRC} that the average delay per hop is
expressed as
\begin{eqnarray}
l_{hop}&=&(\mathbb{E}[\eta]N+\mathbb{E}[m_e+m_n])T_s+n_TTp.\label{eq:l_hop_GeRaF}
\end{eqnarray}
\indent The expected number of hops traversed before reaching the
destination as function of $R$ and $\rho$ is denoted by $q$. It may
be derived using \cite{GeRaF.multihop.zorzi} ((8) and (19)). As
such, the end-to-end energy and delay are $qE_{hop}$ and $ql_{hop}$
respectively.\\
\indent The forwarding process of GeRaF-PC is quite similar, except
for the fact that the data payload is only transmitted once.

\subsection{BOSS}
BOSS was designed such that packet forwarding may well be picked up
by a node in the NPA. Under such circumstances, authors of BOSS
suggest to use a greedy-face-greedy algorithm \cite{BOSS.Sanchez}.
In such a case, we assume that the remaining distance to the
destination stays unchanged. This is indeed a sub-accurate
assumption. However, it
constrains the complexity of the subsequent analysis.\\
\indent Nodes in BOSS compute the value of the response time based
on the subarea they lie within. Nodes offering the best progress
will have the shortest response time. To reduce the probability of
collisions, a random variable is added to the response time. The
variable is uniformly-distributed over the interval $[0,x]$. As a
result, this enhances the granularity at which time is slotted: the
duration of one control slot for BOSS
is $T_c/x$ while for GeRaF is as large as $2T_c$.\\
\indent The forwarding process in BOSS can be conceived to consist
of 4 distinct stages:
\begin{enumerate}
\item Empty cycles \item Cycles where forwarding is picked up from NPA \item Cycles with collisions \item Successful round
\end{enumerate}
The relay selection and packet forwarding process in BOSS is
captured in detail in Table \ref{table:BOSS}. The probability that
forwarding is picked up by a node in NPA is equivalent to the
probability that there exists no awaken nodes in PPA and at least
one awaken node in NPA. As such
\begin{equation}
p_{NPA}=e^{-\zeta\epsilon\rho\pi R^2}(e^{-(1-\zeta)\epsilon\rho\pi
R^2}),
\end{equation}
where $\zeta$ is the ratio of the PPA to the entire coverage area.
The probability that forwarding takes place from PPA is therefore
$p_{PPA}=1-p_{NPA}$. Expressions for $\mathbb{E}[m_e]$ and
$\mathbb{E}[m_n]$ are again found in \cite{GeRaF.latency.zorzi}((3)
and (4)). We also need to derive an expression for the probability
of collision denoted here as $p_c$. This probability can be derived
in light of \cite{OMR.Bader} (7). Given the number of contending
nodes is $n$, then \cite{OMR.Bader} (7) provides an expression for
the probability that the first $j$ slots are not resolvable, i.e.
they carry colliding messages. In our case, we are rather interested
in the event of having a collision after a series of ``empty''
slots. The probability of having $j-1$ empty slots from a pool of
$x$ slots is hence $\left(\frac{x-j+1}{x}\right)^n$. When we are
left with only $x-j+1$ slots, then using \cite{OMR.Bader} (6) and
(7), the probability that the first slot will be non-resolvable is
$1-\left(\frac{x-j}{x-j+1}\right)^{n-1}$, $n\geq 1$. Consequently,
the probability of collision given $n$ is
\begin{equation}
p_{c|n}=1-\sum_{j=1}^{x-1}\left(\frac{x-j}{x-j+1}\right)^{n-1},
n\geq 1.
\end{equation}
The probability mass function (pmf) for the number of nodes existing
in a given forwarding subarea is given by
\begin{equation}
p_n(n|m_e)=\frac{(\epsilon\rho a_{m_e})^n}{n!}e^{-\epsilon\rho
a_{m_e}}.
\end{equation}
With $p_{m_e}(m_e)$ readily available from
\cite{GeRaF.latency.zorzi}(3), we are able now to compute $p_n(n)$
by averaging over $m_e$. We can then compute
$p_c=\sum_{n}p_{c|n}p_n(n)$. The number of cycles with collisions
before a successful round is denoted by $\overline{\eta}$ and
follows a geometric distribution such that
$p_{\acute{\eta}}(\acute{\eta})={p_c}^{\acute{\eta}}(1-p_c)$.
Consequently,
\begin{equation}
\mathbb{E}[\acute{\eta}]=\frac{1-p_c}{p_c}.
\end{equation}
In light of Table \ref{table:BOSS}, the energy consumed in
forwarding a packet per hop can be expressed by summing up all
energy terms (as we have done in the case of GeRaF-MRC). On the
other hand, the delay per hop is given by
\begin{eqnarray}
l_{hop}&=&\mathbb{E}[\eta](T_p+2xNT_s)+p_{NPA}(T_p+(2xN+1)T_s)\nonumber\\
&+ &
p_{PPA}(\mathbb{E}[\acute{\eta}](T_p+xNT_s)+T_p+(\mathbb{E}[m_e]+2)T_s).\nonumber\\
\end{eqnarray}
For a communication range equal to $R$, the average number of hops
is again derived using \cite{GeRaF.multihop.zorzi}((8) and (19)),
such that $E_{e2e}=qE_{hop}$ and $l_{e2e}=ql_{hop}$.

\subsection{RTS/CTS/DATA-opt}
Whereas GeRaF-PC and GeRaF-MRC mandate the use of a CONTINUE
message, it is assumed that RTS/CTS/DATA-opt does not. As mentioned
before, RTS/CTS/DATA-opt is a hypothetical geo-routing protocol
which is able to successfully forward a packet using the minimum
number of message transactions. This is illustrated in Table
\ref{table:opt}. Energy consumed per hop is derived in light of
Table \ref{table:opt} by summing up all energy terms as done before
for BOSS, GeRaF-MRC, and GeRaF-PC. The delay per hop however is
given by
\begin{equation}
l_{hop}=(2N\mathbb{E}[\eta]+2\mathbb{E}[m_e]+3)T_c+T_p.
\end{equation}
Bringing RTS/CTS/DATA-opt into the picture greatly aids in
understanding the performance limits of GeRaF-PC, GeRaF-MRC, and
BOSS. Consequently, we introduce here a composite metric of
performance relative to RTS/CTS/DATA-opt. The metric incorporates
both energy and delay and is defined as
\begin{equation}
C_{e2e}=\varphi\frac{E_{e2e,P}}{E_{e2e,{opt}}}+(1-\varphi)\frac{l_{e2e,P}}{l_{e2e,{opt}}},\label{eq:comp_cost}
\end{equation}
where the subscript $P$ corresponds to any of the protocols
GeRaF-MRC, GeRaF-PC, or BOSS, and $\varphi$ is a weighting
parameter.


\begin{table*}[t]
\caption{Relay Selection and Packet Forwarding Process in GeRaF-MRC}
\begin{center}
\begin{tabular}{|llllll|}
  \hline\hline
  \multicolumn{6}{|c|}{}\\

  \multicolumn{6}{|c|}{\textbf{$\eta$ empty cycles}}\\

\multicolumn{6}{|c|}{}\\
  \hline\hline

  \textbf{Activity}, $y$ & \textbf{Duration}, $t_y$ & \textbf{Node(s)} & \textbf{Avg. count}, $n_y$ & \textbf{Task} & \textbf{Power Factor}, $pf_y$\\
   \hline\hline

1 & $T_c$   &   sender  &   $1$     &   transmit RTS&
$P_{t_{C}}$ \\

\hline

2 & $T_c+(N-1)T_s$     &   sender  &   $1$     & listen and activate
& $P_{Rx}+P_{t_{BT}}$\\
& & & & BT while listening &\\
\hline

   \hline\hline
   \multicolumn{6}{|c|}{}\\

   \multicolumn{6}{|c|}{\textbf{non-empty cycle}}\\

   \multicolumn{6}{|c|}{}\\
   \hline\hline

   \textbf{Activity}, $y$ & \textbf{Duration}, $t_y$ & \textbf{Node(s)} & \textbf{Avg. count}, $n_y$ & \textbf{Task} & \textbf{Power Factor}, $pf_y$\\

    \hline\hline

3 & $T_c$   &   sender  &   $1$     &   transmit RTS    &
$P_{t_{C}}$ \\
\hline

4 & $T_c$ &       awaken nodes in  PPA     & $(\zeta\pi
R^2-\sum_{i=1}^{m_e}a_i) \epsilon\rho$       &       receive RTS,
&       $P_{Rx}+P_{t_{BT}}$\\
& & & & activate BT while receiving &\\
\hline

5 & $m_eT_s$        &       awaken nodes in  PPA  & $(\zeta\pi
R^2-\sum_{i=1}^{m_e}a_i) \epsilon\rho$       & listen in
anticipation of a CTS,        &
$P_{Rx}+P_{t_{BT}}$\\
& & & & activate BT while listening & \\
\hline

6 & $m_eT_s$        &      sender  & $1$       & transmit CONTINUE
message &
$\frac{1}{2}P_{t_C}$\\
& & & & in the 1st half of the slot & \\
\hline

7  &  $m_eT_s$   &      sender  & $1$       & wait for a CTS
response &
$\frac{1}{2}P_{t_C}$\\
& & & & in the 2nd half of the slot, & \\
& & & & activate BT while listening & \\
\hline

8     &    $m_nT_s$        &      relays in the       &        at least $2$         &     transmit CTS response       &    $\frac{1}{2}P_{t_C}$      \\
      &                    &      $(N-m_e-1)$th       &                             &     in the 1st half of the slot,&                              \\
      &                    &      forwarding subarea  &                             &     activate BT while listening &                              \\

\hline

9  &  $m_nT_s$ &   sender      &       $1$     &       detect
colliding
    responses       &
    $\frac{1}{2}(P_{t_C}+P_{Rx}+P_{t_{BT}})$\\
& & & & in the 1st half of the slot, & \\
& & & & activate BT while listening, & \\
& & & & transmit CONTINUE message  & \\
& & & & in the 2nd half of the slot & \\
\hline

10 & $T_c$       &       successful relay        &       $1$ &
transmit CTS        &       $P_{t_{C}}$\\
\hline

11 & $T_c$  &   sender      &       $1$     &       receive CTS, activate BT &      $P_{Rx}+P_{t_{BT}}$\\
\hline

12 & $T_c$       &       sender      &       $1$     &       inform successful relay  &        $P_{t_{C}}$\\
\hline

13 & $T_c$ &       successful relay        &       $1$     & receive
selection message,        &       $P_{Rx}+P_{t_{BT}}$\\
& &       &       &       activate BT meanwhile       &\\
\hline

\hline\hline
   \multicolumn{6}{|c|}{}\\

   \multicolumn{6}{|c|}{\textbf{$n_T$ data packet transmissions}}\\

   \multicolumn{6}{|c|}{}\\
   \hline\hline

   \textbf{Activity}, $y$ & \textbf{Duration}, $t_y$ & \textbf{Node(s)} & \textbf{Avg. count}, $n_y$ & \textbf{Task} & \textbf{Power Factor}, $pf_y$\\

    \hline\hline

14 & $T_p$       &       sender      &       $1$     &
transmit
data payload        &       $P_{t_D}$\\
\hline

15  &    $T_p$  &       selected relay        &       $1$ &
        receive packet      &       $P_{Rx}+P_{t_{BT}}$\\
\hline

 16 & $T_c$       &       selected relay      &       $1$
& transmit
ACK/NACK message        &       $P_{t_C}$\\
\hline

17  &   $T_c$  &       sender      &       $1$     & receive ACK/NACK message,        &       $P_{Rx}+P_{t_{BT}}$\\
   &     &       &       &       activate BT while receiving     &\\
\hline

\hline\hline

\end{tabular}
\end{center}
 \label{table:GeRaF-MRC}
\end{table*}


\begin{table*}[t]
\caption{Relay Selection and Packet Forwarding Process in BOSS}
\begin{center}
\begin{tabular}{|llllll|}
  \hline\hline
  \multicolumn{6}{|c|}{}\\

  \multicolumn{6}{|c|}{\textbf{$\eta$ Empty Cycles}}\\

\multicolumn{6}{|c|}{}\\
  \hline\hline

  \textbf{Activity}, $y$ & \textbf{Duration}, $t_y$ & \textbf{Node(s)} & \textbf{Avg. count}, $n_y$ & \textbf{Task} & \textbf{Power Factor}, $pf_y$\\
   \hline\hline

1& $T_p$   &   sender  &   $1$     &   transmit RTS, payload
included &
$P_{t_{D}}$ \\
\hline

2& $2xNT_s$     &   sender  &   $1$     & listen and activate BT &
$P_{Rx}+P_{t_{BT}}$\\
\hline

\hline\hline
   \multicolumn{6}{|c|}{}\\

   \multicolumn{6}{|c|}{\textbf{$j$ NPA Cycles}}\\
   \multicolumn{6}{|c|}{(in case forwarding takes place from the Negative Progress Area)}\\

   \multicolumn{6}{|c|}{}\\
   \hline\hline

    \textbf{Activity}, $y$ & \textbf{Duration}, $t_y$ & \textbf{Node(s)} & \textbf{Avg. count}, $n_y$ & \textbf{Task} & \textbf{Power Factor}, $pf_y$\\

    \hline\hline

3 & $T_p$   &   sender  &   $1$     &   transmit RTS    &
$P_{t_{D}}$ \\
\hline

4 & $T_p$ &       awaken nodes       &
$\epsilon\rho\sum_{i=N+1}^{2N}a_i$ & receive RTS,
&       $P_{Rx}+P_{t_{BT}}$\\
& &in  NPA & & activate BT while receiving& \\
\hline

5 & $2xNT_s$        &       awaken nodes   &
$\epsilon\rho\sum_{i=N+1}^{2N}a_i$ & transmit CTS messages, &
$\frac{P_{t_C}}{2xN}$\\
& &    in  NPA   &       &       each in its corresponding slot      &\\
\hline

6 & $2xNT_s$    &       sender        &       $1$     &
receive CTS messages,         &       $P_{Rx}+P_{t_{BT}}$\\
& &       &       &       activate BT meanwhile       &\\
\hline

7 & $T_s$       &       sender      &       $1$     &       select relay  &        $P_{t_{C}}$\\
\hline

8 & $T_s$ &       successful relay        &       $1$     & receive
selection message,        &       $P_{Rx}+P_{t_{BT}}$\\
& &       &       &       activate BT meanwhile       &\\
\hline

  \hline\hline
  \multicolumn{6}{|c|}{}\\

  \multicolumn{6}{|c|}{\textbf{$\acute{\eta}$ Cycles with Collisions}}\\
  \multicolumn{6}{|c|}{(in case forwarding takes place from the Positive Progress Area)}\\

  \multicolumn{6}{|c|}{}\\
  \hline\hline

   \textbf{Activity}, $y$ & \textbf{Duration}, $t_y$ & \textbf{Node(s)} & \textbf{Avg. count}, $n_y$ & \textbf{Task} & \textbf{Power Factor}, $pf_y$\\

9 & $T_p$   &   sender  &   $1$     &   transmit RTS    &
$P_{t_{D}}$ \\
\hline

10 & $T_p$ &       awaken nodes in       & $(\pi
R^2-\sum_{i=1}^{m_e}a_i) \epsilon\rho$ &       receive RTS,
&       $P_{Rx}+P_{t_{BT}}$\\
& & NPA $\cup$ PPA & &activate
BT while receiving & \\
\hline

11 & $m_eT_s$        &       awaken nodes in   & $(\pi
R^2-\sum_{i=1}^{m_e}a_i) \epsilon\rho+1$       & listen in
anticipation of a CTS,        &
$P_{Rx}+P_{t_{BT}}$\\
& & NPA $\cup$ PPA plus
sender & & activate BT while listening & \\
\hline

12 & $T_s$       &       colliding relays        &       at least
$2$ &
send CTS on the same slot       &       $P_{t_{C}}$\\
\hline

13 & $T_s$ &       sender      &       $1$     &       attempt to
receive the
CTS,      &        $P_{Rx}+P_{t_{BT}}$\\
& & & & activate BT meanwhile &\\
\hline

14 & $(xN-m_e-1)T_s$     &       colliding relays        & at least
$2$ & listen hoping for ACK from sender,  &
$P_{Rx}+P_{t_{BT}}$\\
& & (assuming other   & & activate BT
meanwhile & \\
& & nodes have already  & & & \\
& & dropped off) & & & \\
\hline

   \hline\hline
   \multicolumn{6}{|c|}{}\\

   \multicolumn{6}{|c|}{\textbf{Successful Round}}\\
   \multicolumn{6}{|c|}{(in case forwarding takes place from the Positive Progress Area)}\\

   \multicolumn{6}{|c|}{}\\
   \hline\hline

   \textbf{Activity}, $y$ & \textbf{Duration}, $t_y$ & \textbf{Node(s)} & \textbf{Avg. count}, $n_y$ & \textbf{Task} & \textbf{Power Factor}, $pf_y$\\

    \hline\hline

15 & $T_p$   &   sender  &   $1$     &   transmit RTS    &
$P_{t_{D}}$ \\
\hline

16 & $T_p$ &       awaken nodes in     & $(\pi R^2-$       & receive
RTS,
activate &       $P_{Rx}+P_{t_{BT}}$\\
& &  NPA $\cup$ PPA & $\sum_{i=1}^{m_e}a_i) \epsilon\rho$ &BT while receiving & \\
\hline

17 & $m_eT_s$        &       awaken nodes in   & $(\pi R^2-$ &
listen in anticipation of a CTS,        &
$P_{Rx}+P_{t_{BT}}$\\
& & NPA $\cup$ PPA plus
sender & $-\sum_{i=1}^{m_e}a_i) \epsilon\rho+1$ & activate BT while listening & \\
\hline

18 & $T_s$       &       successful relay        &       $1$     &
transmit CTS        &       $P_{t_{C}}$\\
\hline

19 & $T_s$ &   sender      &       $1$     &       receive CTS, activate BT &      $P_{Rx}+P_{t_{BT}}$\\
\hline

20 & $T_s$       &       sender      &       $1$     &       inform successful relay  &        $P_{t_{C}}$\\
\hline

21 &   $Ts$&     successful relay        &       $1$     & receive
selection message,        &       $P_{Rx}+P_{t_{BT}}$\\
& &       &       &       activate BT meanwhile       &\\
\hline

\hline\hline

\end{tabular}
\end{center}
 \label{table:BOSS}
\end{table*}


\begin{table*}[t]
\caption{Optimal 3-way handshake (RTS/CTS/DATA-opt)}
\begin{center}
\begin{tabular}{|llllll|}
  \hline\hline
  \multicolumn{6}{|c|}{}\\

  \multicolumn{6}{|c|}{\textbf{$\eta$ Empty Cycles}}\\

\multicolumn{6}{|c|}{}\\
  \hline\hline

\textbf{Activity}, $y$ & \textbf{Duration}, $t_y$ & \textbf{Node(s)} & \textbf{Avg. count}, $n_y$ & \textbf{Task} & \textbf{Power Factor}, $pf_y$\\
 \hline\hline

1 & $T_c$   &   sender  &   $1$     &   transmit RTS &
$P_{t_{D}}$ \\
\hline

2 & $NT_c$     &   sender  &   $1$     & listen and activate BT    &
$P_{Rx}+P_{t_{BT}}$\\
\hline

   \hline\hline
   \multicolumn{6}{|c|}{}\\

   \multicolumn{6}{|c|}{\textbf{Successful Round}}\\

   \multicolumn{6}{|c|}{}\\
   \hline\hline

   \textbf{Activity}, $y$ & \textbf{Duration}, $t_y$ & \textbf{Node(s)} & \textbf{Avg. count}, $n_y$ & \textbf{Task} & \textbf{Power Factor}, $pf_y$\\

    \hline\hline

3 & $T_c$   &   sender  &   $1$     &   transmit RTS    &
$P_{t_{D}}$ \\
\hline

4 & $T_c$ &      awaken nodes     & $(\zeta\pi R^2-$       & receive
RTS,
&       $P_{Rx}+P_{t_{BT}}$\\
& & in PPA  & $\sum_{i=1}^{m_e}a_i) \epsilon\rho$ & activate BT while receiving &\\
\hline

5 & $m_eT_c$        &       awaken nodes in  & $(\zeta\pi R^2-$ &
listen in anticipation of a CTS,        &
$P_{Rx}+P_{t_{BT}}$\\
& & PPA plus sender  & $\sum_{i=1}^{m_e}a_i) \epsilon\rho+1$ & activate BT while listening & \\
\hline

6 & $T_c$       &       successful relay        &       $1$     &
transmit CTS        &       $P_{t_{C}}$\\
\hline

7 & $T_c$ &   sender      &       $1$     &       receive CTS, activate BT &      $P_{Rx}+P_{t_{BT}}$\\
\hline

8 & $T_c$       &       sender      &       $1$     &       inform successful relay  &        $P_{t_{C}}$\\
\hline

9 & $T_c$ &       successful relay        &       $1$     & receive
selection message,        &       $P_{Rx}+P_{t_{BT}}$\\
& &       &       &       activate BT meanwhile       &\\
\hline

10 & $T_p$       &       sender      &       $1$     &
transmit
packet      &       $P_{t_D}$\\
\hline

 11 & $T_p$ &      relay       &       $1$     &   receive
packet &
$P_{Rx}+P{t_{BT}}$\\
\hline

 \hline\hline

\end{tabular}
\end{center}
 \label{table:opt}
\end{table*}


\section{Application Scenarios}\label{sec:eval}
\begin{figure*}[t]
\begin{center}
\subfigure[End-to-end delay.]{
\includegraphics[width=5.85cm]{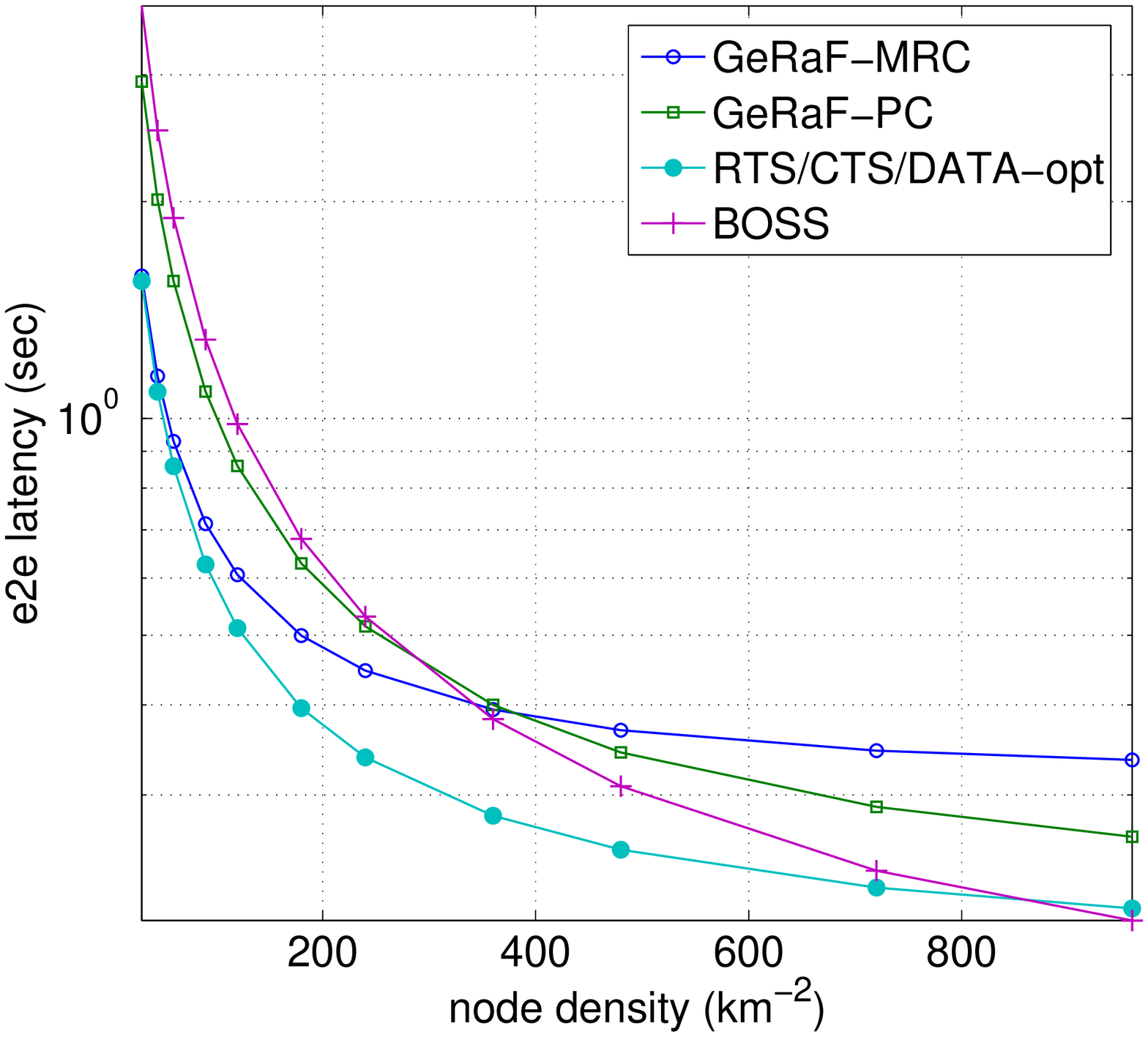}\label{fig:rho_latency}}
\subfigure[Energy consumed to transport one bit from source to
destination.]{
\includegraphics[width=5.85cm]{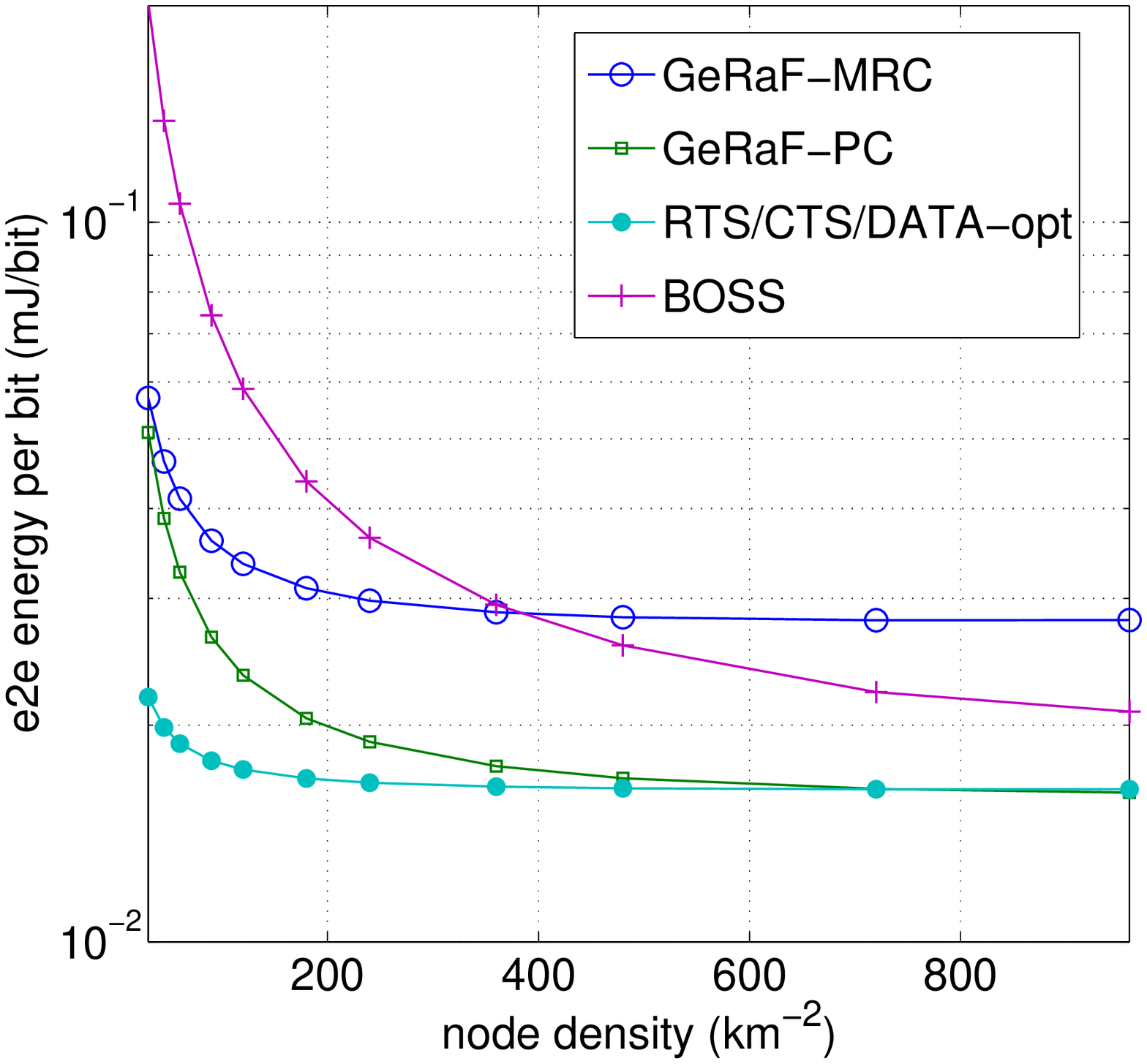}\label{fig:rho_energyperbit}}
\subfigure[End-to-end composite metric, with $\varphi=0.5$]{
\includegraphics[width=5.85cm]{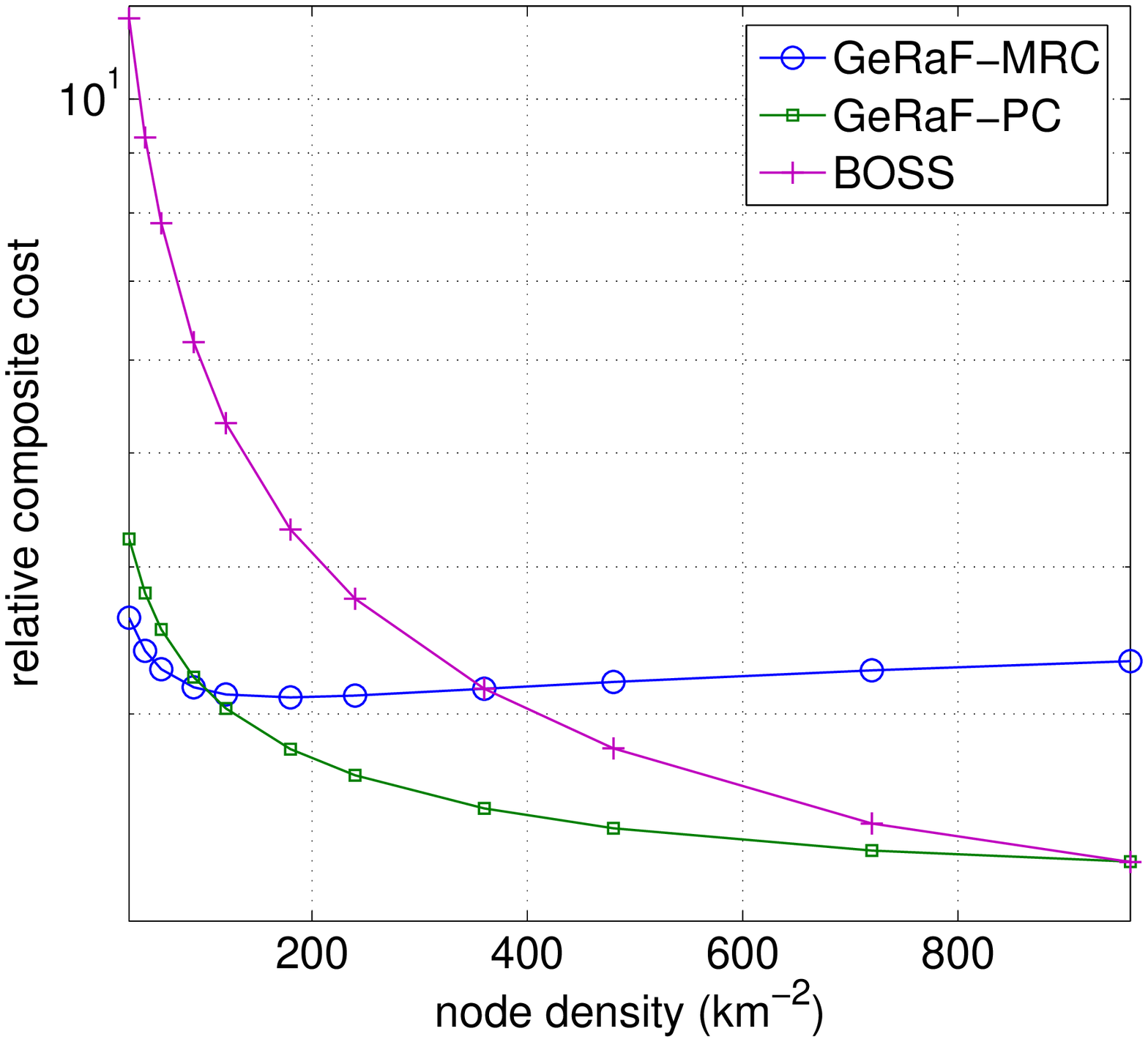}\label{fig:rho_relcost}}
\end{center}
\caption{End-to-end performance of GeRaF-PC, GeRaF-MRC, and BOSS as
function of the network node density.} \label{fig:rho}
\end{figure*}

With the ability to compute end-to-end energy and delay, we are able
now to study the performance of geo-routing protocols over a wide
range of parameters and application scenarios. Before doing so, we
look more closely at how these protocols perform as function of node
density which is indeed an intrinsic characteristic of any wireless
network. Figure \ref{fig:rho} shows the end-to-end performance as
function of the underlying node density.\\
\indent There are a few noteworthy observations here. First, we note
how well GeRaF-MRC performs in terms of delay at low node densities
(Figure \ref{fig:rho_latency}). In fact, at very low densities
GeRaF-MRC clearly outperforms BOSS and GeRaF-PC and approaches
optimality (represented by RTS/CTS/DATA-opt). The reason behind this
is the fact that GeRaF-MRC transmits both data and control packets
at the maximum available power, thus achieving larger hop distances.
As a consequence, the number of hops required is less. Furthermore,
sparse node densities accompanied with shorter hop distances in case
of GeRaF-PC induce many more empty cycles. However, at higher node
densities, BOSS starts to demonstrate better delay performance. The
strategy of introducing more granularity to the time domain in case
of BOSS indeed pays off at higher node densities. On the flip side
of the coin, excellent delay performance for BOSS comes at the
expense of more energy consumption as per Figure
\ref{fig:rho_energyperbit}. As shown in the figure, it is GeRaF-PC
which offers unparalleled performance in terms of energy. GeRaF-PC
reaches nearly optimal energy performance starting from medium node
densities. With such an excellent energy performance, GeRaF-PC is
able to offer the lowest composite cost metric as shown in Figure
\ref{fig:rho_relcost}.

\indent In this section, we further evaluate the performance of
beaconless geo-routing protocols from the perspectives of four
specific application scenarios:
\begin{enumerate}
\item VANET scenarios in which traffic and road safety information are exchanged between vehicles.
\item Rescue field operations in which members of a
rescue team communicate voice and video data with each other.
\item Transmission of meter readings and usage patterns in Smart Utility Networks (SUNs).
\item Environmental or process monitoring applications such as monitoring air
quality in urban areas or temperature variations in an industrial
process.
\end{enumerate}
The analytical framework developed in this paper comes handy in
identifying the application areas or scenarios where each protocol
is better positioned. These four application scenarios are discussed
in greater detail next.

\subsection{Vehicular Ad Hoc Networks}
\begin{figure}[t]
\begin{center}
\subfigure[As function of node density]{
\includegraphics[width=7cm]{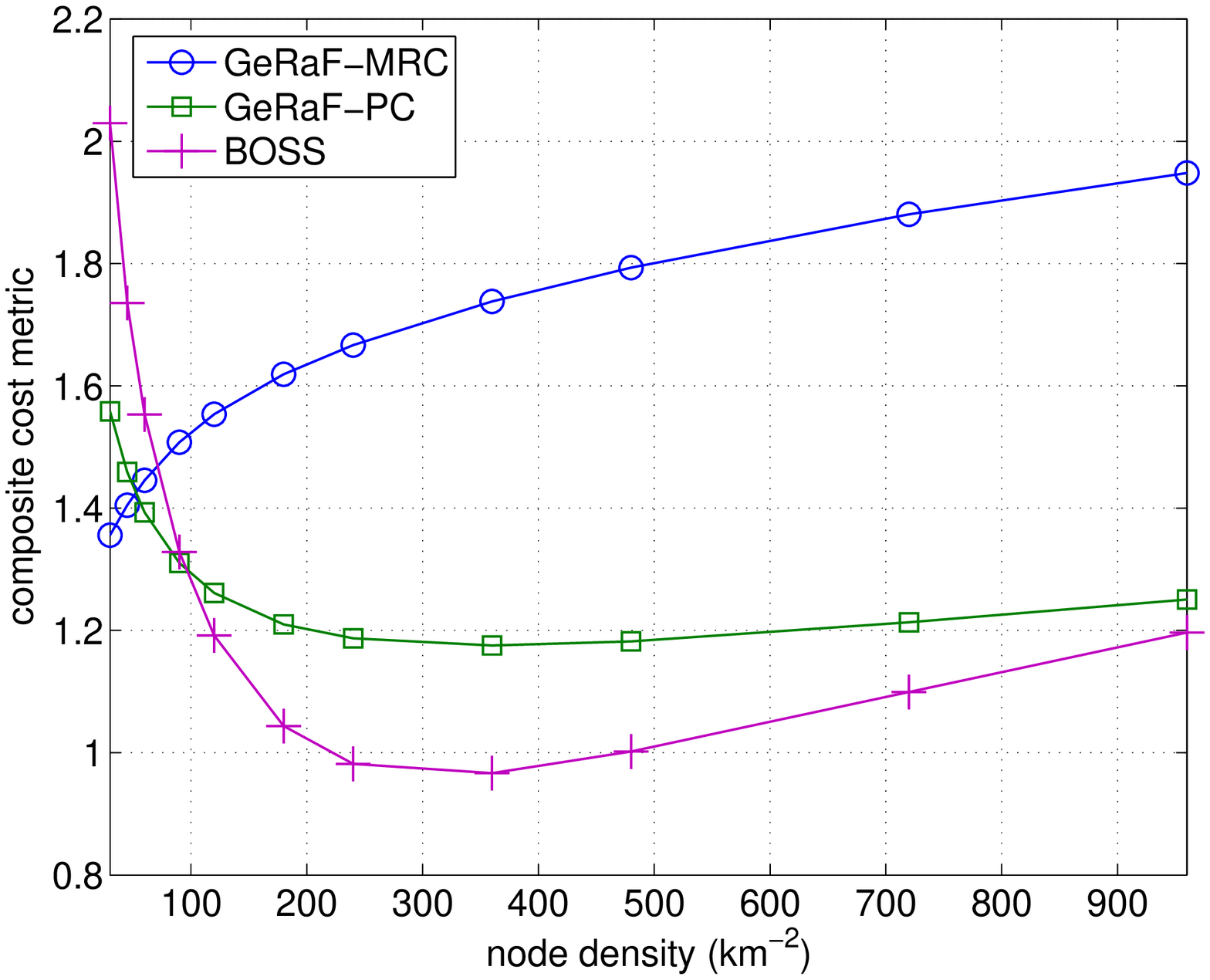}\label{fig:VANET_rho}}
\subfigure[As function of maximum available transmit power]{
\includegraphics[width=6.7cm]{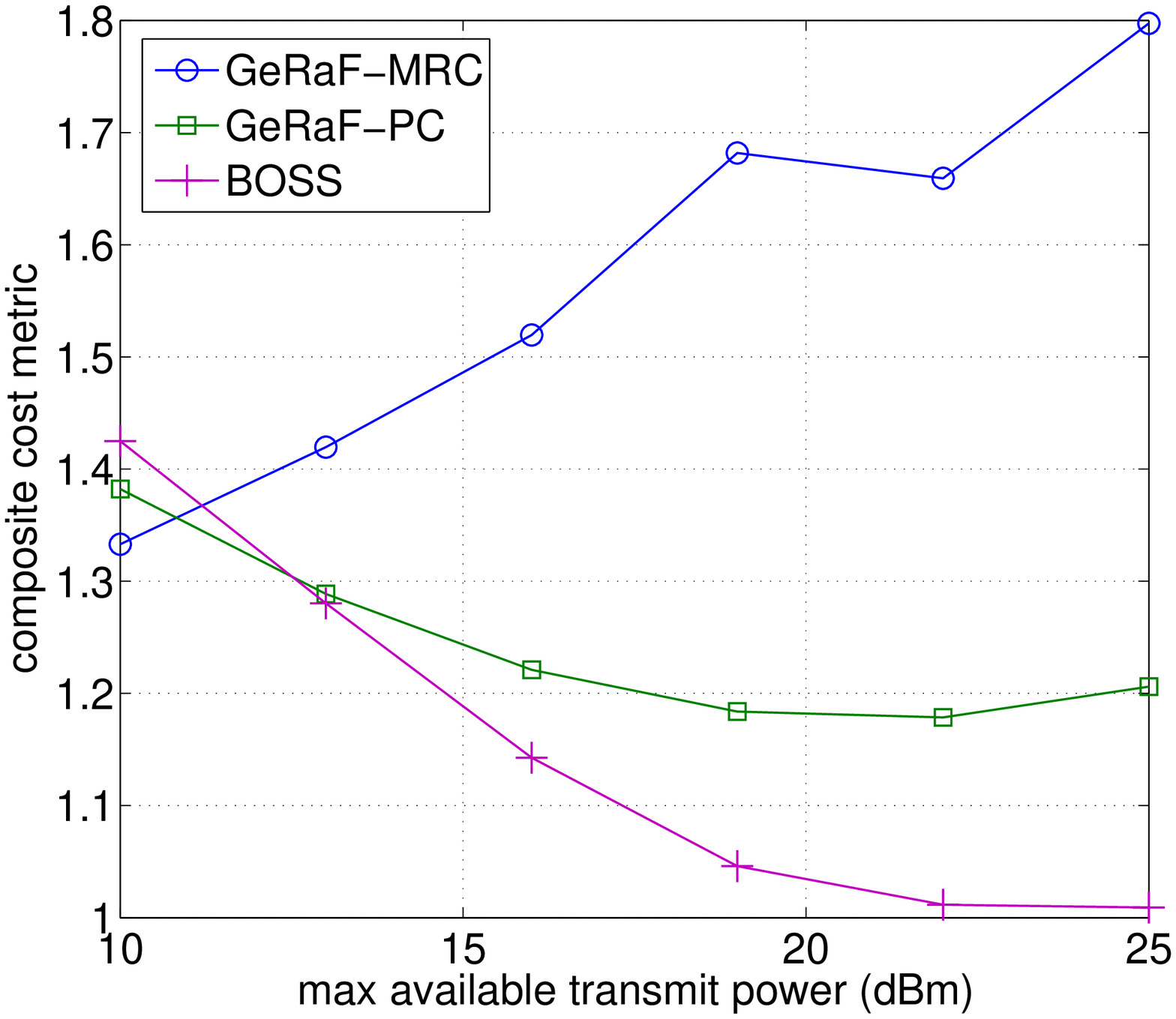}\label{fig:VANET_Pt}}
\end{center}
\caption{Composite performance metric for a VANET application
scenario.} \label{fig:VANET}
\end{figure}

The wireless channel in a VANET is going to be frequency selective
as explained in Section \ref{section:PHY}. A node in a VANET is
privileged with access to relatively abundant energy. Therefore,
nodes can transmit at a higher radio power level. Furthermore,
energy consumption is not a primary concern. As such, we can grant
end-to-end delay performance more attention by setting the weighting
factor $\varphi=0.2$ in (\ref{eq:comp_cost}). Moreover, the
availability of energy eliminates the need for sleeping in VANETs,
i.e. $\epsilon=1$. In urban and dense urban scenarios, node density
is high. With
$\epsilon=1$, the node density is virtually even higher.\\
\indent Figure \ref{fig:VANET} captures performance results in terms
of the composite performance metric for a VANET scenario. It is
quite evident that BOSS is the protocol of choice in this case,
particularly at medium node densities. In highly dense networks,
GeRaF-PC might be able to keep up with BOSS (Figure
\ref{fig:VANET_rho}). However, increasing the transmit power gives
grounds back to BOSS as shown in Figure \ref{fig:VANET_Pt}. For
instance, when the transmit power is set at greater than 22 dBm,
performance of BOSS comes very close to the limit.

\subsection{Rescue Field Networks}
\begin{figure*}[t]
\begin{center}
\subfigure[Absolute end-to-end delay.]{
\includegraphics[width=8cm]{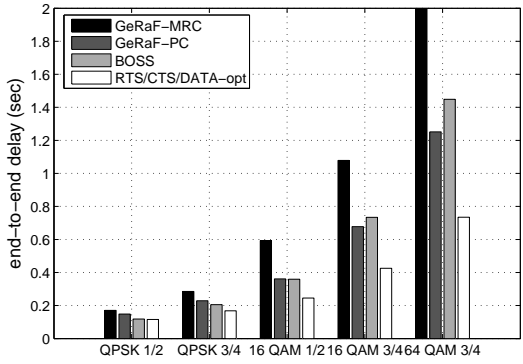}\label{fig:rescue_delay}}
\subfigure[End-to-end delay normalized by the number of bits
transported per packet.]{
\includegraphics[width=8cm]{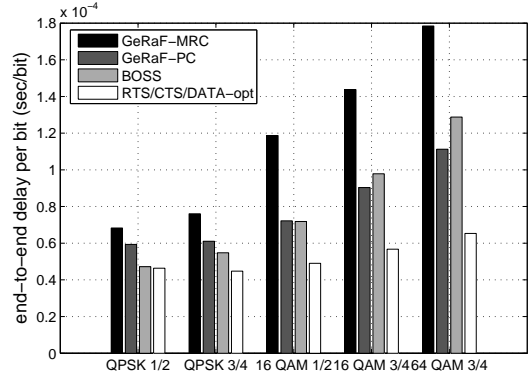}\label{fig:rescue_delay_perbit}}
\subfigure[End-to-end energy consumed in transferring one bit from
the source to the destination.]{
\includegraphics[width=8cm]{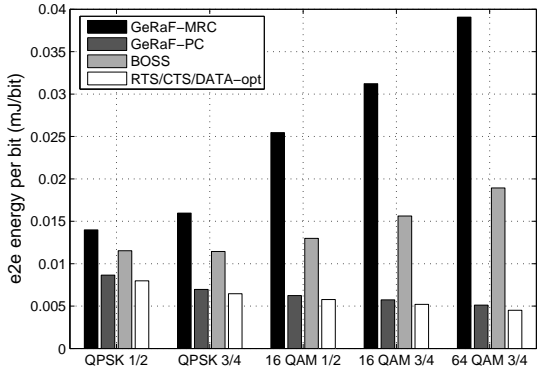}\label{fig:rescue_energy_perbit}}
\subfigure[End-to-end composite cost metric with $\varphi=0.6$.]{
\includegraphics[width=8cm]{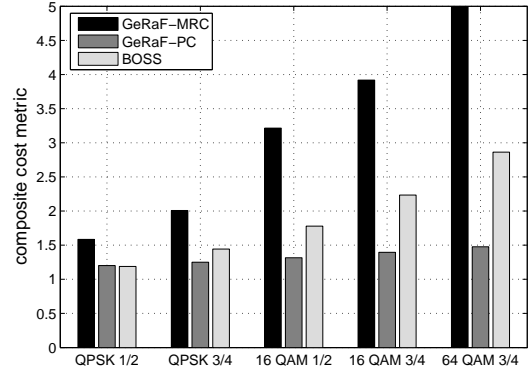}\label{fig:rescue_comp}}
\end{center}
\caption{End-to-end performance results for rescue field networking
applications.} \label{fig:rescue}
\end{figure*}

Fire fighters and rescue teams would highly benefit from the
availability of voice and video communications during their
operations. Higher data rates would be required under such
circumstances. We assume that the PHY layer is capable of meeting
this requirement in the sense that is able to support higher oreders
of modulation and coding schemes (MCS)\footnote{The IEEE
802.15.4-2006 standard actually only supports 250 kbps.
Nevertheless, for the sake of this study we assume higher rates are
achievable by means of adapting the MCS.}. On the other hand,
shorter packet lengths would be typically the case here since voice
and video are the primary type of data. The duration of the packet
size is set here at $T_p=10$ ms.\\
\indent Delay performance is of large importance for this
application scenario. Energy comes at second priority since users
may have the chance to recharge the batteries of their devices upon
the completion of each mission. Nevertheless, energy consumption is
still a concern especially for operations of long durations.
Consequently, we set $\varphi=0.6$. Needless to mention here that
$\epsilon=1$, i.e. nodes do not sleep due to the risk and human
safety factors associated
with such applications.\\
\indent Figure \ref{fig:rescue} shows end-to-end performance of
GeRaF-MRC, GeRaF-PC, and BOSS for 5 different modulation and coding
schemes: QPSK $\frac{1}{2}$, QPSK $\frac{3}{4}$, 16 QAM
$\frac{1}{2}$, 16 QAM $\frac{3}{4}$, and 64 QAM $\frac{1}{2}$. We
first note that all protocols suffer from growth in the end-to-end
delay as the MCS rank is upgraded, as shown in Figure
\ref{fig:rescue_delay}. This is intuitive since communication ranges
are reduced due to the increase in the SINR requirement. At the same
time however, the number of bits transported end-to-end also
increases. As such, we normalize the end-to-end delay by the number
of bits as shown in Figure \ref{fig:rescue_delay_perbit} which
conveys a less steep growth in the performance metric as MCS is
upgraded. It is also worthwhile to note from Figure \ref{fig:rescue}
that GeRaF-MRC performs quite poorly at higher MCS ranks. This is
due to the fact that the SINR gap between the control and data
packets grows significantly and thus the number of data packet
transmissions $n_T$ faces a sheer increase. In terms of energy per
bit, it is clear also that BOSS does not perform very well. For
rescue field networks, GeRaF-PC is clearly the best choice since it
consistently offers the lowest relative cost metric as Figure
\ref{fig:rescue_comp} reveals.

\subsection{Smart Utility Networks}
Remote home metering and transfer of usage patterns is one important
aspect of smart utility networks \cite{WSN.smart_cities.Onworld}.
For such an application, energy performance is very important since
users will not be willing to recharge the batteries very often. On
the other hand, delay is substantially insignificant since meter
readings collection may take place only every few weeks. As such,
the focus in a SUN application is to study the energy performance.\\
\indent Sleeping is an essential practice in this application
scenario as it helps save energy. So it is favorable to apply
immense sleeping patterns. The virtual effect of immense sleeping is
a steep decline in node density. An additional property of a SUN is
the requirement for only modest data rates. The performance of
beaconless geo-routing for SUN applications is thus best viewed by
varying the node density as a study parameter. Looking again at
Figure \ref{fig:rho_energyperbit}, it is clear that GeRaF-PC is best
fit to serve remote utility metering applications.

\subsection{Environmental Monitoring Networks}
Environmental and process monitoring applications are very similar
in terms of their requirements to smart utility networks, except for
the fact that delay may hold larger significance. This is mainly
because some level-crossing events may mandate a rapid response,
e.g. process temperature crossing a hazard limit, or liquid level in
tank near to cause spills. Looking back again on Figure
\ref{fig:rho_latency}, GeRaF-MRC is shown to provide the best
end-to-end delay performance. Nevertheless, if energy and delay have
equal importance, GeRaF-MRC has an edge only for very low node
densities as illustrated in Figure \ref{fig:rho_relcost}. As node
density increases, GeRaF-PC becomes a better choice.

\section{Parameter Optimization}\label{sec:opt}
\begin{figure*}[t]
\begin{center}
\subfigure[End-to-end delay performance.]{
\includegraphics[width=9cm]{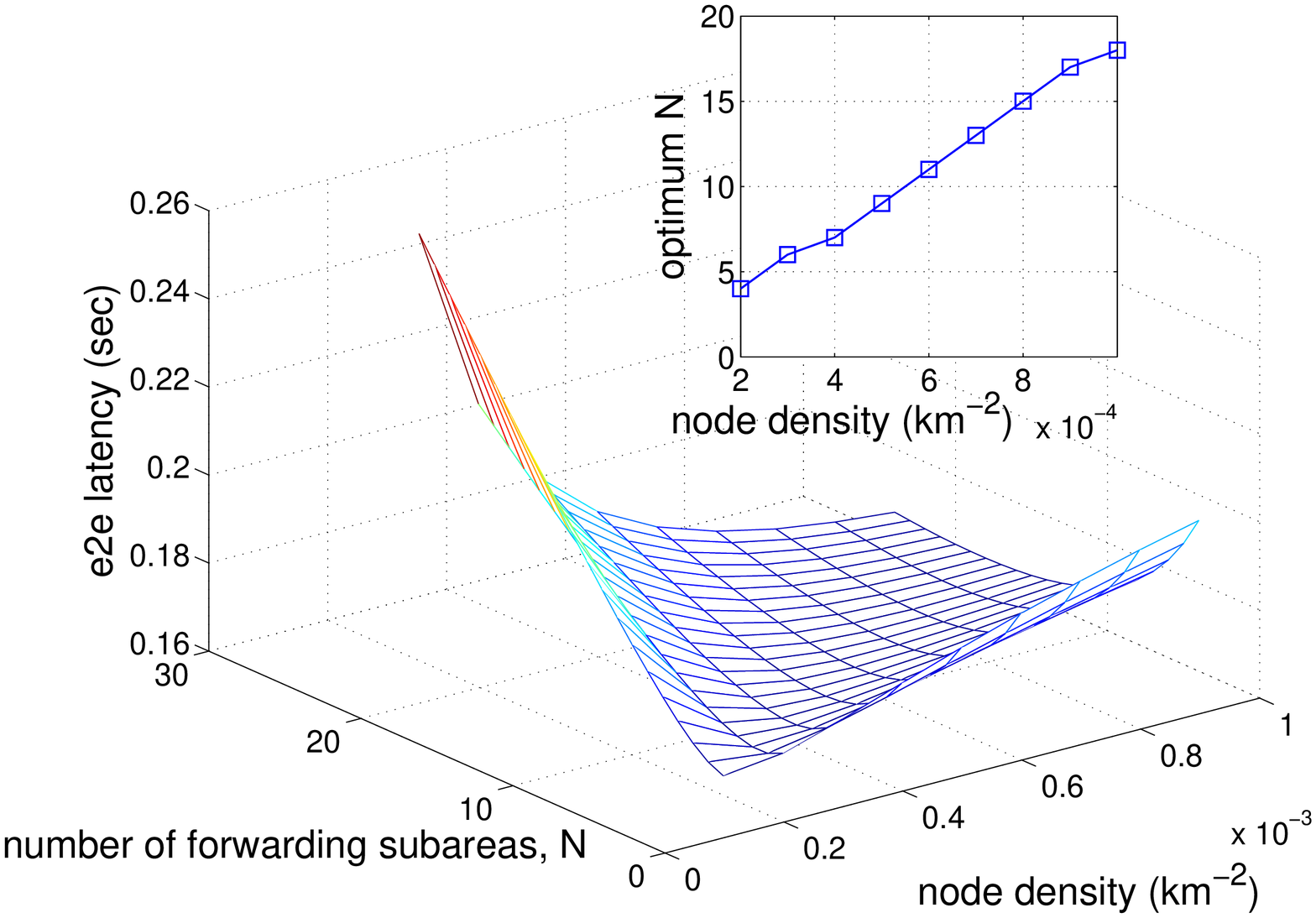}\label{fig:opt_N_latency}}
\subfigure[End-to-end energy performance.]{
\includegraphics[width=9cm]{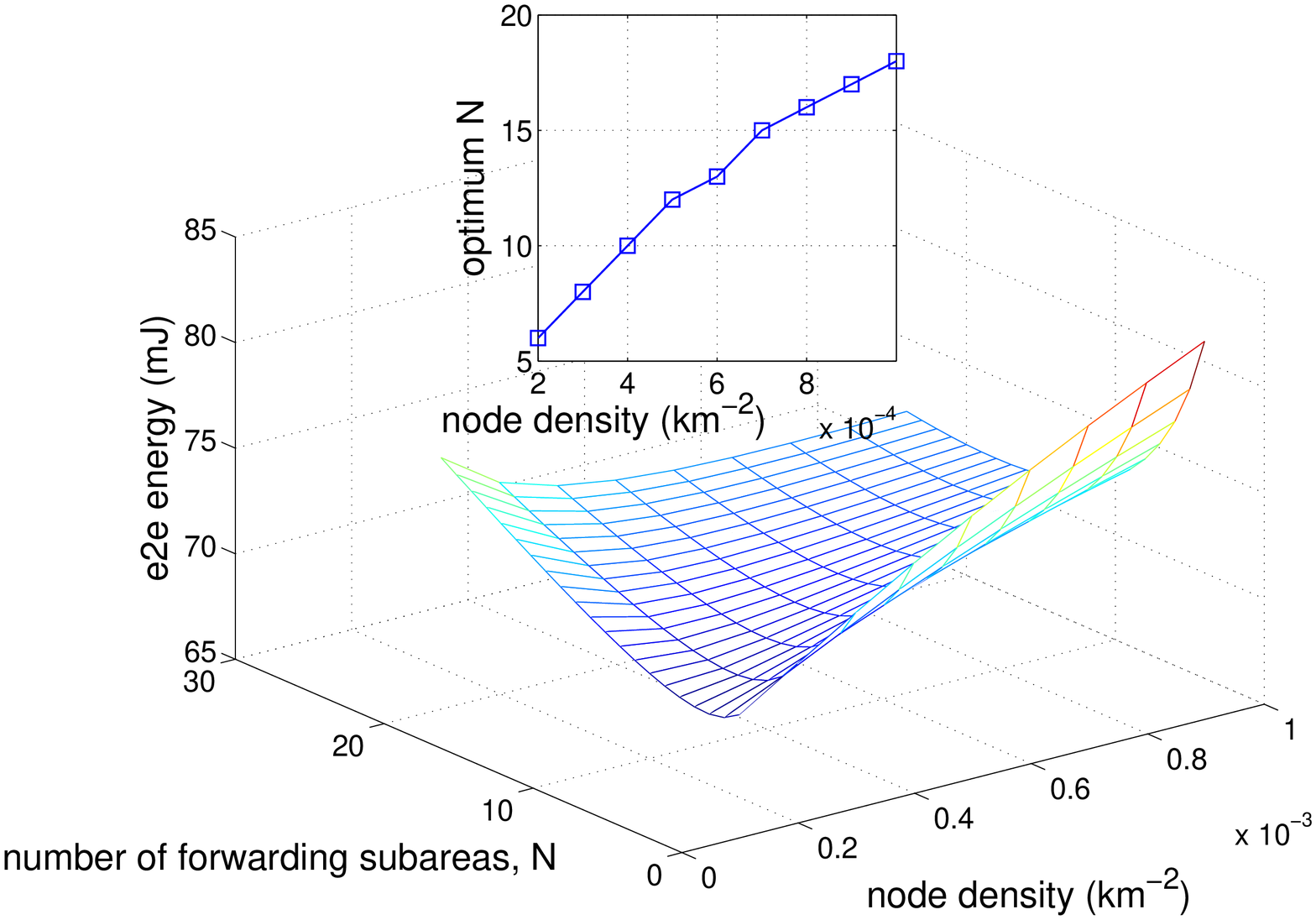}\label{fig:opt_N_energy}}
\subfigure[End-to-end composite cost metric.]{
\includegraphics[width=9cm]{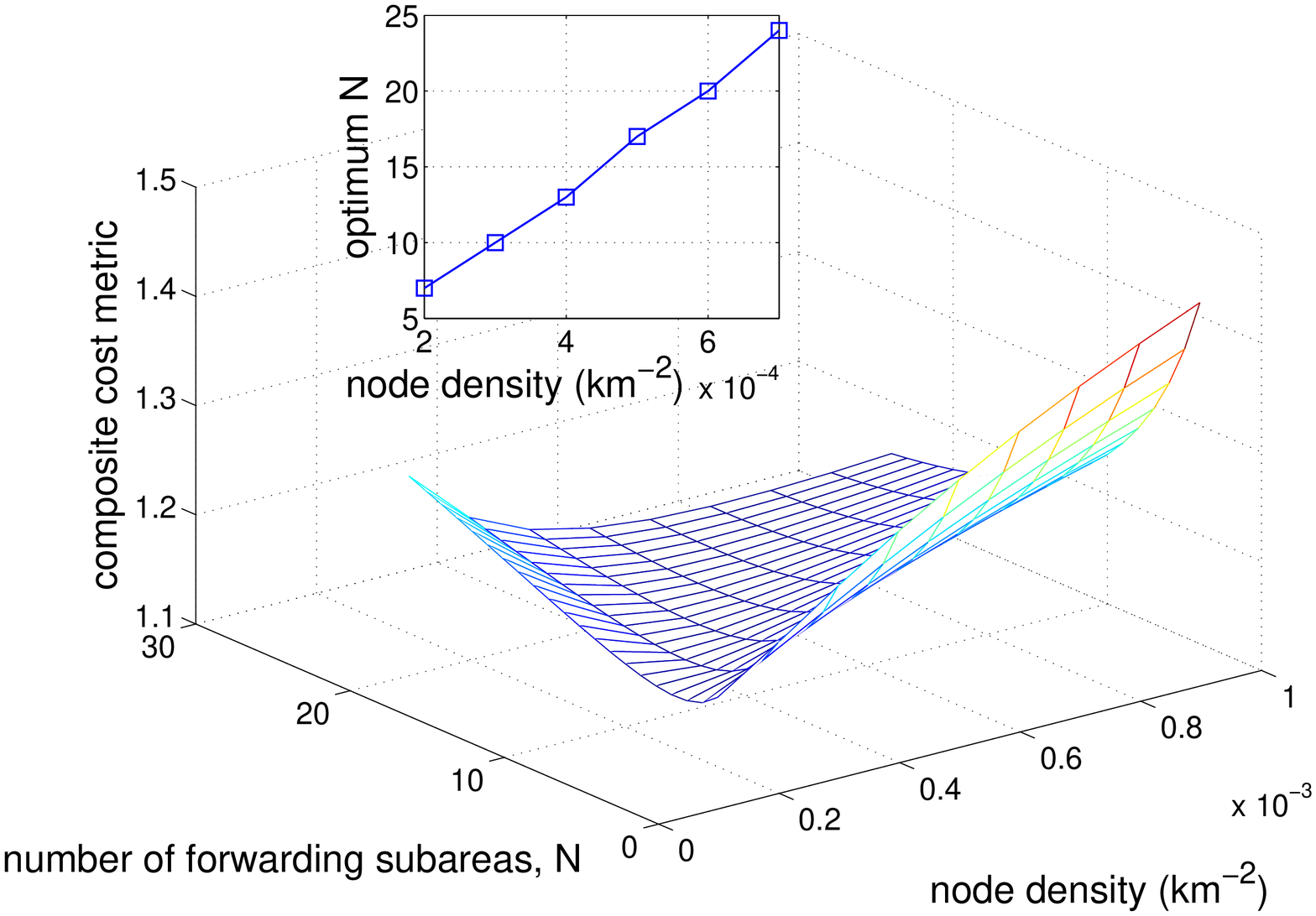}\label{fig:opt_N_compcost}}
\end{center}
\caption{Optimizing over the number of forwarding subareas, $N$.
GeRaF-PC is studied in this specific example. In each subfigure, the
optimum value of $N$ is shown as function of the node density.}
\label{fig:N}
\end{figure*}

\begin{figure}[t]
\begin{center}
\subfigure[End-to-end delay elapsed to transport one bit of
information.]{
\includegraphics[width=5.85cm]{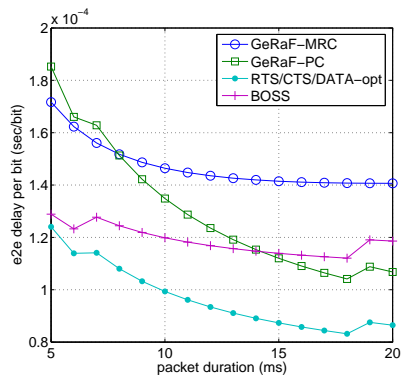}\label{fig:Tp_latencyperbit}}
\subfigure[Energy to transport one bit end-to-end.]{
\includegraphics[width=5.85cm]{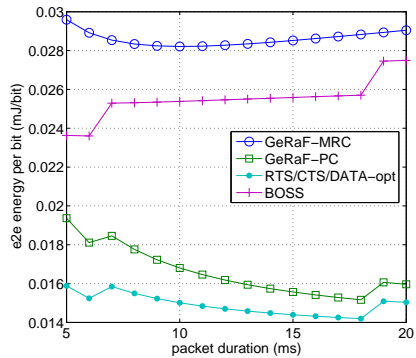}\label{fig:Tp_energyperbit}}
\subfigure[End-to-end composite cost metric.]{
\includegraphics[width=5.85cm]{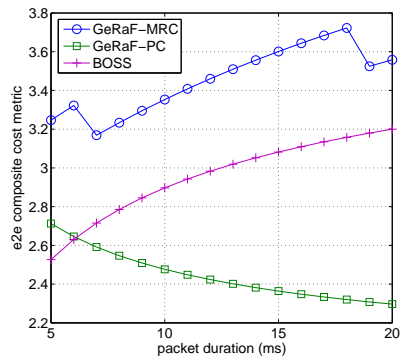}\label{fig:Tp_relcost}}
\end{center}
\caption{Performance of beaconless geo-routing protocols over a
range of packet durations.} \label{fig:Tp}
\end{figure}

We can also utilize the analytical framework developed in this paper
to optimize protocol performance. For instance, we can optimize the
performance of GeRaF-PC over the number of forwarding subareas, $N$.
Figure \ref{fig:N} shows the behavior of end-to-end performance
metrics in response to variations in the $N$. It is clear that the
optimum value of $N$ increases with node density. For considerable
ranges of node densities, the end-to-end performance function
(whether latency, energy, or composite cost) is convex. In other
words, an absolute optimum value for $N$ exists. We further note
that it is preferable at low densities to keep $N$ small. We recall
that increasing the value of $N$ may be needed to reduce the
probability of collisions. Nevertheless, doing that is not necessary
for low densities since the probability of collision is anyway
small. Otherwise, increasing $N$ would only lead to increasing the
delay since it entails a proportional increase in the number of time
slots per
cycle.\\
\indent It is also interesting to study performance as function of
data packet durations. It is important to study performance from
this perspective since some applications place some constraints on
packet lengths. For instance, vector-based data applications (i.e.
voice and video) typically require short packets
\cite{WMSN.survey.Akyildiz(Elsevier)}. So the question here is: how
well do GeRaF-PC, GeRaF-MRC, and BOSS perform for short packet
lengths? Figure \ref{fig:Tp} provides the answer. It is quite
intuitive to expect the end-to-end delay to increase as the packet
duration $T_p$ increases. For the sake of a more meaningful insight,
we have instead normalized the delay by the number of bits when
constructing the plots of Figure \ref{fig:Tp}. In other words, we
seek the amount of end-to-end delay incurred in transporting a
single information bit from source to destination. The same
normalization philosophy is applied when considering end-to-end
energy.\\
From Figure \ref{fig:Tp_latencyperbit}, we note that the normalized
end-to-end delay for all three protocols drops as the packet length
increases. However, it is worthwhile also to note that they all
diverge away from optimality. From the perspective of normalized
energy performance - Figure \ref{fig:Tp_energyperbit} - we see see
clearly that GeRaF-PC slightly improves as $T_p$ increases,
GeRaF-MRC features an almost steady performance, while BOSS inclines
more energy per bit. Taking energy and delay jointly into
consideration - Figure \ref{fig:Tp_relcost} - shows that only
GeRaF-PC enjoys better performance at larger values of $T_p$. So if
the designer has some flexibility in choosing the packet duration,
then there is no doubt that short packet lengths are generally
favorable for GeRaF-PC but not for GeRaF-MRC and BOSS.

\section{Conclusions}\label{sec:conclusions}
In this paper, we have developed an analytical framework for the
end-to-end performance of two prominent beaconless geo-routing
protocols: GeRaF and BOSS. In doing so, we have utilized a practical
packet detection model in order to account for the discrepancy
between the communication ranges of the data and control packets. In
line with this practical model, two new versions of GeRaF have been
devised. The first one applies a power headroom on data packets and
is labeled as GeRaF-PC. The second utilizes MRC techniques to bridge
the gap between control and data packets and is tagged as GeRaF-MRC.
Using the analytical framework developed herewith, four different
application scenarios have been studied: VANETs, rescue field
networks, smart utility networks, and environmental monitoring. It
is shown in this paper that BOSS is optimal for VANETs since it
offers the best delay performance. On the other hand, GeRaF-PC is
the best choice for rescue field networks, as it is able to cope
well with video communication requirements. GeRaF-PC is also very
well-positioned for smart utility applications. For environmental
monitoring applications, GeRaF-MRC is often the protocol with best
performance. Finally, we have exemplified how the analytical
framework can be used to optimize some of the protocol parameters.
For instance, we have shown that the optimum number of forwarding
subareas increases with node density in the case of GeRaF-PC.

\bibliographystyle{IEEEbib11}
\bibliography{bibfileshort}
\balance

\end{document}